\documentclass[aps,prl,reprint,superscriptaddress]{revtex4-1}
\usepackage{graphicx}
\usepackage{dcolumn}
\usepackage{bm}
\usepackage{amsmath}

\usepackage{amsfonts}
\usepackage{textcomp} 

\usepackage{verbatim}
\usepackage {ulem}

\usepackage{color}
\usepackage{array}
\usepackage{makecell}

\newcolumntype{C}{>{\centering\arraybackslash$}m{2cm}<{$}}

\begin{document}

\title{Separation of the two-magnon scattering contribution to damping for the determination of the spin mixing conductance}

\author{A. Conca}

\email{conca@physik.uni-kl.de}

\author{S.~Keller}

\author{M.~R.~Schweizer}

\author{E.~Th.~Papaioannou}

\author{B.~Hillebrands}

\affiliation{Fachbereich Physik and Landesforschungszentrum OPTIMAS, Technische Universit\"at
Kaiserslautern, 67663 Kaiserslautern, Germany}

\date{\today}

\begin{abstract}
We present angle dependent measurements of the damping properties of  epitaxial Fe layers with MgO, Al and Pt capping layers. Based on the preferential distribution of lattice defects following the crystal symmetry, we make use of a model of the defect density to separate the contribution of two-magnon scattering to the damping from the isotropic contribution originating in the spin pumping effect, the viscous Gilbert damping and the magnetic proximity effect. The separation of the two-magnon contribution, which depends strongly on the defect density, allows for the measurement of  a value of the effective  spin mixing conductance which is closer to the value  exclusively due to spin pumping. The influence of the  defect density for bilayers systems due to the different capping layers and to the unavoidable spread in defect density from sample to sample is thus removed.  This shows the potential of studying spin pumping phenomena in fully ordered systems in which this separation is possible, contrary to polycrystalline or amorphous metallic thin films.

\end{abstract}

\maketitle

\section{Introduction}

In bilayers systems formed by a ferromagnetic (FM) layer  in contact with a metallic non-magnetic (NM) one, a pure spin current can be generated and injected in the latter when the ferromagnetic resonance is excited. Typically, a microwave magnetic field is used for this purpose. The whole process is commonly referred to as spin pumping  \cite{tser,tser2}. If the non-magnetic layer is formed by a heavy metal with large spin-orbit coupling (Pt, Ta or similar), the spin current can be detected by using the inverse spin Hall effect (ISHE) for conversion into a charge current.



Since the  spin current leaving the magnetic layer carries away angular momentum from the magnetization precession, it represents an additional loss channel for the magnetic system and consequently causes an increase in the measured Gilbert damping parameter $\alpha$ \cite{tser}:

\begin{equation} \label{deltasp}
\Delta\alpha_{\rm sp}= \frac{\gamma\hbar}{4\pi M_s \:d_{\rm FM}}g^{\uparrow\downarrow}
\end{equation}
where $g^{\uparrow\downarrow}$ is the real part of the spin mixing conductance which  is  controlling the magnitude of the generated spin current and $\gamma$ is the gyromagnetic ratio.

This expression is only valid for sufficiently thick  NM layers where no reflection of the spin current takes place at the film surface or interface with other materials, i.e. no spin current is flowing back into the magnetic layer. In principle, it allows the estimation of $g^{\uparrow\downarrow}$ by measuring the increase in damping compared to the intrinsic value. However, to perform this measurement is not straightforward. If the estimation of $g^{\uparrow\downarrow}$ for a FM/Pt system is needed, ideally one should  measure the effective Gilbert damping parameter $\alpha_{\rm 0}$ for a single standing magnetic layer acting as a reference sample with no losses due to spin pumping and repeat the same after depositing a thick Pt layer. However, most of the common ferromagnetic materials, with exception of the magnetic insulators like YIG, will change its properties due to oxidation processes. Therefore, a capping layer is required and one has to find an appropriate one, in the sense that its introduction must not modify the damping properties of the magnetic layer. Examples in the literature show that this is far to be a trivial task \cite{ana,cfa2018,nata}. In addition to this, the emergence of a finite magnetic polarization in Pt in contact with a ferromagnetic layers has an impact on damping which further hinders the estimation of $g^{\uparrow\downarrow}$ \cite{lu,suzuki,wilhelm,fept,ana,caminale,gosh,lack}.

For the reference layers, the most convenient candidates as capping material are oxides like MgO, for which it has been proven that they are able to block the flow of spin current and therefore to deactivate spin pumping \cite{eid,mihalceanu,mosendz}, or metals with weak spin-orbit interaction like Al or Ru. But even for these cases, it has been shown that an increase of damping not related to spin pumping is possible. Ruiz \textit{et al.} show for instance that a MgO capping layer increases strongly the damping in permalloy while this is not the case for Al capping layer \cite{ana}. The reason has nothing to do with the metallic character of the capping layer since the increase for Ru is even larger than with MgO. The same work \cite{ana} already provides a hint for a possible reason since the increase of damping roughly scales  with the value of the interface perpendicular anisotropy constant $K^{\perp}_{\rm S}$. Theoretical works \cite{usov} show that the counterplay between the demagnetizing field responsible for the in-plane orientation of the magnetization and the  perpendicular anisotropy field can induce inhomogeneous magnetization states for certain field strengths combinations which are responsible for an increased damping. In this sense, this effect has been also adduced to explain the damping thickness dependence in Co$_2$FeAl/MgO systems \cite{cfa2018}.

\begin{figure*}[t]
    \includegraphics[width=1.0\textwidth]{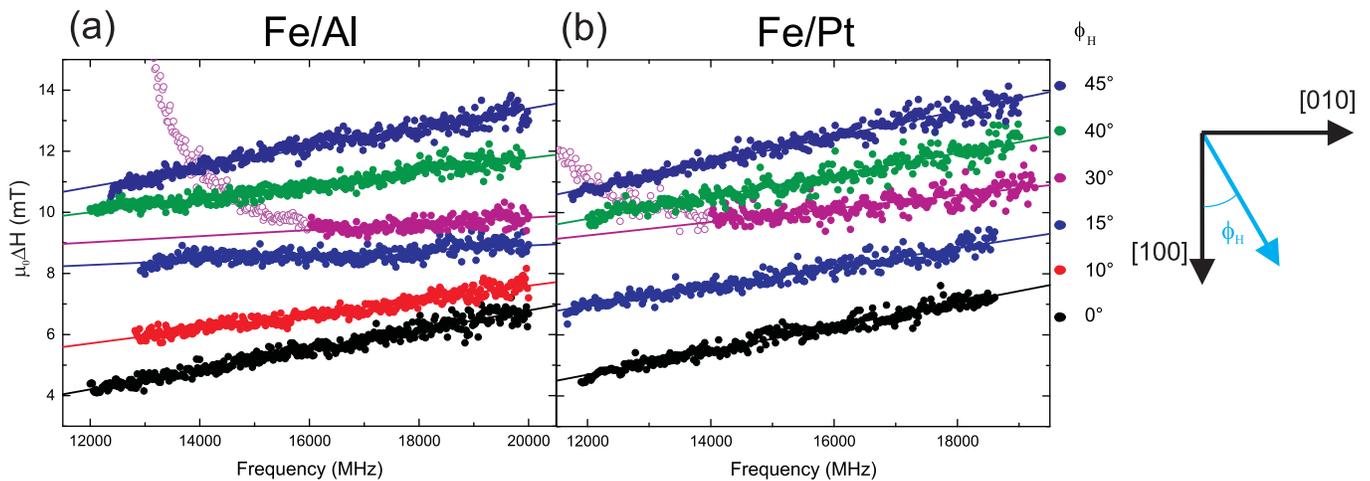}
	  \caption{\label{deltah}(Color online) Dependence of the FMR linewidth on the frequency for different orientations $\phi_H$ of the external magnetic field with respect to the [100] crystallographic axis of Fe for  (a) Fe/Al and  (b) Fe/Pt systems. The lines correspond to a linear fit to extract the effective damping parameter $\alpha_{\rm eff}$. For $\phi =30^{\circ}$ a strong non-linearity due to magnetic dragging is observed.  For visibility reasons, each data set is shifted vertically by 1.25 mT with respect the previous one.}
\end{figure*}

Here we present angle dependent measurements of the damping properties of  epitaxial Fe layers with MgO, Al and Pt capping layers. Fully epitaxial systems constitute a perfect ordered model with almost ideal and well defined interfaces. Here, we will show that the angle dependence of damping allows for a measurement of the strength of the two-magnon scattering and of its contribution to the effective  damping parameter. With the separation of this contribution we access the increase of damping caused only by spin pumping and magnetic proximity effect and to an estimation of $g^{\uparrow\downarrow}$ without the contamination of defects effects.

\section{Experimental details}

The samples were deposited by e-beam evaporation on MgO(100) substrates in a molecular beam epitaxy (MBE) chamber with a base pressure P$\rm _b=5\times10^{-10}$~mbar. A set of Fe/Pt bilayers  with fixed Fe thickness (12~nm) and varying Pt thickness were prepared. Additional reference samples, where Pt is substituted by MgO or Al, have also been prepared.  The Fe and Pt films were grown with a deposition rate of 0.05\AA /s.  The samples were deposited  with a substrate temperature of $300^{\circ}$C and subsequently annealed at the same temperature.

The  characterization by X-ray diffractometry (XRD) (presented elsewhere \cite{keller}) shows that the Fe/Pt bilayers are fully epitaxial with the Fe unit cell rotated by $45^{\circ}$ with respect to the MgO substrate unit cell and with Pt  rotated again $45^{\circ}$ with respect to Fe. In the case of Fe/Al,  epitaxial growth of the upper layer could not be achieved.

The dynamic properties and material parameters were studied by measuring the ferromagnetic resonance using a strip-line vector network analyzer (VNA-FMR). For this, the samples were placed facing the strip-line and the $\tilde{\rm S}_{12}$ transmission parameter was recorded.

\section{Results and discussion}

Figures~\ref{deltah}  shows the dependence of the measured FMR line width $\Delta H$ on the frequency for the reference layer with Al capping (a) and a Fe/Pt system (b). The data is shown for different orientation of the external static magnetic field varying from $\phi_H=0^{\circ}$ ([100], easy axis)  to $\phi_H=45^{\circ}$  ([110], hard axis). For visibility reasons, each data set is shifted vertically by 1.25~mT with respect to the previous one. 

As commented before, the choice of capping layer can have a large influence on the linewidth and effective damping of the magnetic layer, even for light metals.  The magnetic proximity effect (MPE) in the case of Pt also contributes to an increase on damping, \cite{fept,ana,caminale,gosh,lack} which additionally challenges the measurement of the contribution from the spin pumping. Taking into account all these considerations, the effective increase on damping when comparing a reference system and a system with a heavy metal can be separated as follows:

 \begin{equation} \label{alpha}
\alpha_{\rm eff}= \alpha_{\rm 0}+\alpha_{\rm mpe}+\alpha_{\rm sp}+\alpha_{\rm i}.
\end{equation}

Here $\alpha_{\rm 0}$ is the intrinsic damping parameter which can be defined as characteristic of the material under investigation (growth conditions however may influence it strongly) and it is the sum of the losses by two-magnon scattering and by energy transfer to the phonon system. $\alpha_{\rm mpe}$ is the contribution due to the dynamic coupling between the ordered spins in Pt due to the MPE and the  magnetization in the magnetic layer. $\alpha_{\rm sp}$ is the result of the losses by the spin current generated in the ferromagnetic layer by the precession of the magnetization and that flows into the Pt layer (spin pumping). The last term $\alpha_{\rm i}$ summarizes the increase of damping due to other interfacial effects such as interface PMA as commented above, spin memory loss \cite{khasa} or isotropic scattering at interface defects \cite{ing}.

\begin{figure}[t]
    \includegraphics[width=1.0\columnwidth]{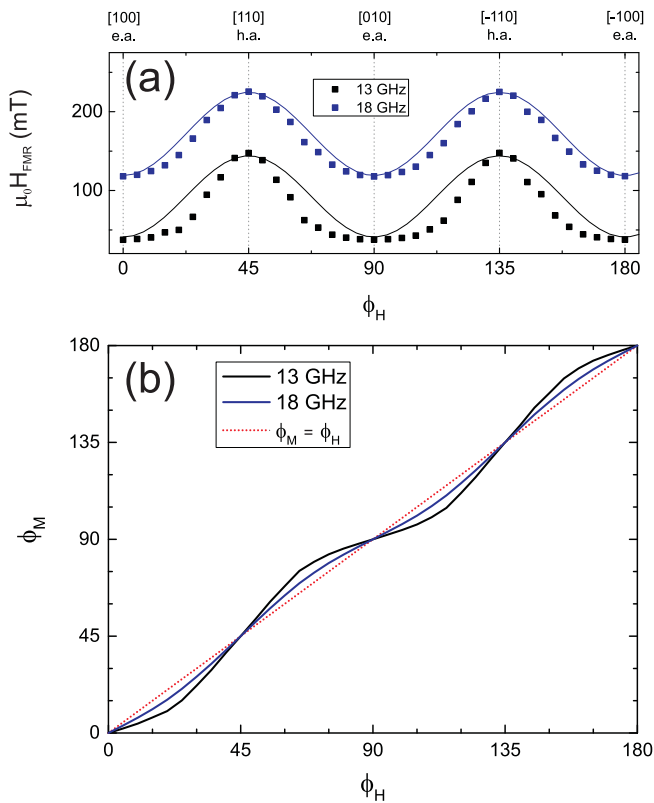}
	  \caption{\label{nocollinear}(Color online) (a) Dependence of the FMR resonance field $H_{\rm FMR}$ on the in-plane direction of the static magnetic field for two values of the resonant frequency. (b) Dependence of the in-plane angle of the magnetization vector $\phi_M$ on the external field direction $\phi_H$. Both angles are measured relative to the [100] axis.  The dotted line represents the case of perfect collinearity between magnetization and external field.}
\end{figure}

Several efforts have been made in order to separate some of the contributions to $\alpha_{\rm eff}$. In a recent work with CoFeB/Pt \cite{lack} we were able to separate $\alpha_{\rm mpe}$ due to the dependence on the Pt thickness. As already reported by Caminale \textit{et al.} \cite{caminale},   a linear  Pt thickness dependence of the spin-current absorption in spin-sink layers  exhibiting MPE and of $\alpha_{\rm mpe}$ is expected \cite{gosh}. 
A detailed vector network analyzer FMR study has also been recently reported to separate the different contributions in NiFe/Pt systems \cite{berger}. 

The term $\alpha_{\rm 0}$ is a result of two contributions \cite{lenz}. One is the pure Gilbert damping, which is of viscous nature and generates a dissipation of energy and angular momentum to the lattice. The second one is the transfer to spin-wave modes with $k\neq 0$ from the FMR mode via two-magnon scattering. For a pure Gilbert-like viscous damping the linewidth dependence on the frequency is purely linear:

\begin{equation} \label{gilbert}
\mu_0\Delta H= \mu_0\Delta H_0 + \frac{4 \pi \alpha f}{\gamma}.
\end{equation}
Here, $\Delta H_0$ is the inhomogeneous broadening and is related to film quality.

The lines in Figs.~\ref{deltah} (a) and (b) are a fit to this expression. It has to be mentioned that although a viscous damping generates a linear dependence, on the contrary it is not possible to assume that the observation of a linear behavior proves that only viscous damping is present. The reason for that is that two-magnon scattering can mimic also a linear dependence \cite{lenz,aria-mills,belmeguenai2013}. For both samples, and for the MgO capped sample not shown here, for $\phi =30^{\circ}$  a strongly non linear behavior with a large increase in linewidth values for smaller frequencies is observed. For this reason, the hollow points in Fig.~\ref{deltah} have been  excluded from the fit. The non-linearity at low frequencies cannot be explained by viscous damping and  it is caused by magnetic dragging. The magnetic dragging effect describes the increase of the linewidth of precessing magnetic layers with large magnetic crystaline anisotropy due to the non-collinearity of the magnetization and the external magnetic field. In Fig.~\ref{nocollinear} (a), the dependence of the resonance field $H_{\rm FMR}$ on the in-plane direction of the external magnetic field is shown for two fixed frequency values. As a result of the four-fold anisotropy expected from the cubic lattice of Fe and assuming a perfect collinearity between magnetization vector and external field,  $H_{\rm FMR}$ can be modeled as: \cite{heinrich,fept}

\begin{equation} \label{HFMRpolar}
\mu_0H_{\rm FMR}= \mu_0\tilde{H}_{\rm FMR}+\frac{2K_1}{M_s}\text{cos}(4\phi),
\end{equation}
where $K_1$ is the cubic anisotropy constant, $\phi$ the in-plane azimuthal angle and $\tilde{H}_{\rm FMR}$ is the averaged resonance field value. The fraction $\frac {2K_1}{M_s}$ is directly the anisotropy field H$_{\rm B}$. In Fig.~\ref{nocollinear}(a) a deviation from this model is observed  for angles between the hard and easy axis and it is due to magnetic dragging, i.e., the magnetization is not aligned to the external field due to the effect of the anisotropy field. The fact that the deviation from the model in Eq.~\ref{HFMRpolar} is smaller for larger frequencies (i.e. larger applied field) also supports this interpretation.
The same behavior  observed for $\phi =30^{\circ}$ has been also been reported for ultrathin Fe films \cite{chen} or for insulating LSMO films \cite{barsukov2016} and attributed to magnetic dragging. The degree of non-collinearity can be estimated by solving the equilibrium condition for  the angle defining the orientation of the magnetization $\phi_M$ for each value of $\phi_H$:

\begin{equation} \label{equilibrium}
H \text{sin}(\phi_M-\phi_H)+\frac{H_{\rm B}}{4} \text{sin}(4\phi_M )=0,
\end{equation}
where the value for the cubic anisotropy field was taken from \cite{fept}. Fig.~\ref{nocollinear}(b) shows the obtained value of $\phi_M$ for the data shown in Fig.~\ref{nocollinear}(a). The angle between magnetization and magnetic field can be as large as 10$^{\circ}$ for 13~GHz and it is decreased to a maximum around 4.5$^{\circ}$ for 18~GHz. The magnetic dragging effect is largest for  $\phi_H$ between the easy and hard axis and vanishes along the main crystallographic axes.

\begin{figure*}[t]
    \includegraphics[width=1.0\textwidth]{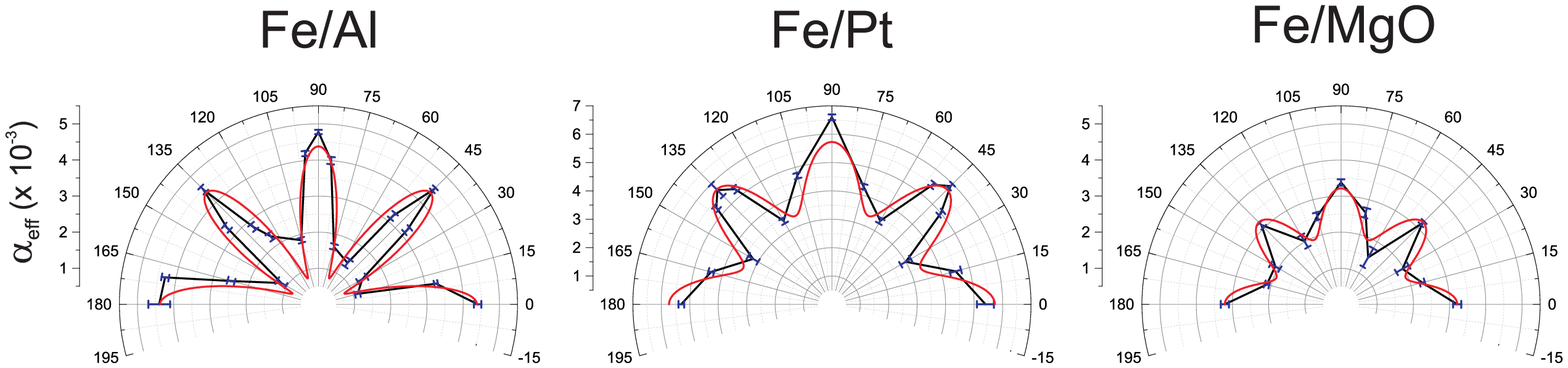}
	  \caption{\label{alpha_angular}(Color online) Angular dependence of the effective damping parameter $\alpha_{\rm eff}$ in the in-plane direction of the static magnetic field $\phi_H$ for (a) Fe/Al, (b) Fe/Pt and (c) Fe/MgO. The red lines are a fit to Eq.~\ref{alphafit}. }
\end{figure*}

Figure~\ref{alpha_angular} shows the value of the effective damping parameter $\alpha_{\rm eff}$ as obtained from the fits in Fig.~\ref{deltah} for the three capping layers. In all of them, an eight-fold symmetry on the in-plane angle $\phi_H$ is observed with maxima along the easy and hard axis of the Fe layers and minima in between. For the Fe/Al and Fe/MgO samples, where spin pumping has no influence,  $\alpha_{\rm eff}= \alpha_0 + \alpha_{\rm i}$ while for the Fe/Pt sample, where both losses through spin pumping and due to the MPE are active, we obtain the situation shown in Eq.~\ref{alpha}. It is remarkable that the different origins of the damping do not change the overall symmetry of the angular dependence. It has though an impact on the absolute values, which are larger for the Fe/Pt sample.

In the literature concerning epitaxial layers, it is possible to find different symmetries for the dependence of the FMR linewidth or the damping parameter on the in-plane field direction. For the Heusler alloy Co$_2$FeAl both four- and eight-fold symmetries for the linewidth have been reported. The situation differs depending on the thickness of the film \cite{belmeguenai2013} and also between different groups \cite{chen2018} pointing out to a role of the growth conditions. For Fe$_3$Si films and Fe/V multilayer systems a four-fold symmetry is reported \cite{zakeri,lenz} and for ultrathin Fe layers, where the role of the interface is strong, a two-fold symmetry of $\alpha_{\rm eff}$ has been measured \cite{chen}. Eight-fold symmetry has been also observed in epitaxial FeSi systems \cite{zakeri,barsukov2010}. In a different work on Fe layers, a decrease on the obtained $\alpha$ value along the intermediate orientation between the two main axis  relative to the one measured along the easy and hard axis  was reported \cite{meckenstock}, pointing to an angular dependence very similar to ours. Concerning insulating systems, two- and four-fold symmetries have been observed in LSMO films \cite{barsukov2016}.

Two-magnon scattering can only occur if scattering centers in form of defects are present. If, as expected, these are present as point lattice defects or dislocation lines along the main crystallographic directions, it is clear that the scattering intensity should reflect the symmetry of the lattice. This fact would for certain explain a four- or eight-fold  anisotropy in damping observed in some on the reports mentioned above and the maxima in $\alpha_{\rm eff}$ for our samples for $\phi=0^{\circ},45^{\circ},90^{\circ},135^{\circ}$. 

Following Zakeri \textit{et al}. and Aria \textit{et al}., the contribution to damping due to two-magnon scattering can be written as \cite{zakeri,aria-mills}:

\begin{equation} \label{2mag}
\alpha_{\rm 2M}=\sum_{\left\langle x_i\right\rangle} \Gamma_{\left\langle x_i\right\rangle} f(\phi_H-\phi_{\left\langle x_i\right\rangle}),
\end{equation}
where $\Gamma_{\left\langle x_i\right\rangle}$ represents the strength of the two-magnon scattering contribution along the in-plane crystallographic direction $\left\langle x_i\right\rangle$. The function $f(\phi_H-\phi_{\left\langle x_i\right\rangle})$ allows for an angle dependent two-magnon contribution to damping with respect to the orientation of the external field $H$ relative to the crystallographic directions $\left\langle x_i\right\rangle$. The physical interpretation of the function $f(\phi_H-\phi_{\left\langle x_i\right\rangle})$ lays in the Fourier transform of the defects in the film \cite{zakeri,woltersdorf}. By using the ansatz  $f(\phi_H-\phi_{\left\langle x_i\right\rangle})=\text{cos}^2(4\phi_H-\phi_{\left\langle x_i\right\rangle})$ we can fit the damping dependence using a simplified version:

 \begin{equation} \label{alphafit}
\alpha_{\rm eff}= \alpha_{\rm iso} + \alpha_{\rm 2M} = \alpha_{\rm iso}+\Gamma_{\rm 2M}\text{cos}^2(4\phi_H-\phi_{[100]})  
\end{equation}
where $\alpha_{\rm iso}$ includes now all the isotropic contributions to damping, i.e. $\alpha_{\rm mpe}$, $\alpha_{\rm sp}$, pure Gilbert damping and potentially isotropic interface contributions from the term $\alpha_{\rm i}$, mainly spin memory loss and interface PMA related effects.

\begin{table}[b]
\renewcommand{\arraystretch}{1.2}
\begin{tabular}{l c c c}
\Xhline{3\arrayrulewidth}
 & $\alpha_{\rm iso}$  & &$\Gamma_{\rm 2M}$    \\
&(10$^{-3}$)& ~~~~~& (10$^{-3})$ \\ 
\Xhline{3\arrayrulewidth}
Fe/Al 		& 0.8 $\pm$ 0.3 		&	&		3.6 $\pm$ 0.4 		 \\
Fe/Pt			& 3.4 $\pm$ 0.3 		&	&		2.4 $\pm$ 0.4	 \\ 
Fe/MgO		& 1.9 $\pm$ 0.1 		&	&		1.3 $\pm$ 0.1	  \\ 

\Xhline{3\arrayrulewidth}
\end{tabular}
\caption{\label{parameters}Isotropic contribution $\alpha_{\rm iso}$ and two-magnon scattering contribution $\Gamma_{\rm 2M}$ to the total effective damping parameter $\alpha_{\rm eff}$. }
\end{table}

The red lines in Fig.~\ref{alpha_angular} show the fit to this model. The obtained parameters are summarized in Table.~\ref{parameters}. A very low value below 1$\times10^{-3}$ is obtained for $\alpha_{\rm iso}$ for the Fe/Al sample. Since $\alpha_{\rm sp, MPE}=0$  is expected and due to the low value we consider that the obtained $\alpha_{\rm iso}$ must be very close to the  value corresponding only to pure viscous Gilbert damping corresponding to high quality Fe. However, strictly speaking, the obtained value is only an upper limit since still other effects might contribute. Concerning 3d metals with no half-metallic character, a very low damping value of 0.7$\times10^{-3}$  has been reported by Lee \textit{et al.} for CoFe \cite{lee2017}. This value is comparable to the $\alpha_{\rm iso}$  measured here for Fe/Al. The fact that the CoFe samples in which the low value was obtained are also fully epitaxial with an exceptionally high crystalline quality explains the similarity in values. The low defect density in CoFe  almost suppresses two-magnon scattering in the CoFe samples and therefore is comparable with our $\alpha_{\rm iso}$ where that contribution is already separated.

For the Fe/MgO sample the value for $\alpha_{\rm iso}$ increases by a factor larger than 2 although also here $\alpha_{\rm sp, MPE}=0$. The main differences between Fe/Al and Fe/MgO are  that the MgO is single crystalline while Al is polycrystalline and the contrast between the metallic character of Al with the insulating oxide. The lattice mismatch between MgO and Fe is around 4\% and introduces therefore a certain degree of stress in the Fe layer which is not present when the capping is polycrystalline Al and which can have an impact on damping. At the same time, since the Gilbert damping is sensitive to the density of states and this one is modified at the interface by the kind of bonds between the Fe atom and the atoms from the capping layer, the simple material difference may also explain the difference. In this sense it is remarkable that the low damping value by Lee \textit{et al.}  commented before is only observed for CoFe with a MgO capping layer and a larger value is measured when MgAl$_2$O$_4$ is used \cite{lee2017}. Our data  confirms the important role of the capping layer on damping observed in other works \cite{ana}.

A further increase in the value of $\alpha_{\rm iso}$ is observed for the Fe/Pt sample where additional losses through spin pumping and MPE are present. Unfortunately the data presented in this paper does not allow to disentangle these two contributions. For this reason, when using Eq.~\ref{deltasp} for the calculation of spin mixing conductance, it makes sense to refer to an effective value $g^{\uparrow\downarrow}_{\rm eff}$ which is at the same time an upper limit for the corresponding value for spin pumping alone. Using the Fe/Al sample as a reference we obtain a value for the spin mixing conductance of $3.7 \pm 0.9\times 10^{19}$m$^{-2}$. This value is lower than the one presented in our previous report \cite{fept} and shows that the value of $g^{\uparrow\downarrow}_{\rm eff}$ can be easily overestimated if the effect of two-magnon scattering on damping is not separated, with the consequent overestimation of the injected spin current and underestimation of the spin Hall angle from the ISHE voltage \cite{keller}. The advantage of using epitaxial magnetic layers is that they allow the separation of the contribution of the two-magnon scattering due to the strong angular dependence and well defined crystallographic directions. The same is not possible in commonly used material as CoFeB or NiFe where the amorphous or polycrystalline nature of the layers blends the scattering dependence on the in-plane angle. 

The parameter $\Gamma_{\rm 2M}$ provides further insight into the origin of total damping in the samples. This parameter is  larger for the Fe/Al sample in comparison to the fully epitaxial bilayers being almost three times larger than for Fe/MgO. As a result, the total damping in the Fe/Al sample is dominated by the two-magnon scattering due also to the low $\alpha_{\rm iso}$ while the same is not true in the other two systems. It has to be taken into account that, since as scattering centers for magnon scattering  the defects at the interfaces play a role, they can be dominant in thin films. From TEM images (presented for instance in \cite{fept}), we can prove the existence of a highly ordered interface in the fully epitaxial samples. Of course, the same is  not true for the case with polycrystalline Al capping. We believe that the dominant role of the interface here is possible, also due to the overall low defect density in the bulk of the Fe layer.

For completeness we want to discuss two additional effects potentially affecting the linewidth and damping.  Due to the spread of internal  and anisotropy field due to mosaicity in the film, there is a contribution to the line broadening which has the following form \cite{zakeri,michael}:

\begin{equation} \label{mosaic}
\Delta H_{mosaic}= \left|  \frac{\partial H_{\rm FMR}}{\partial \phi_H}\right|\Delta \phi_H,
\end{equation}
where $\Delta \phi_H $ is the average spread of the direction of the easy axes in the film plane. From Fig.~\ref{deltah}(c) it is clear that this contribution should increase the linewidth in the region  $\phi=15-30^{\circ}$ and equivalent ones but this is not observed pointing to a weak impact of mosaicity. In any case, the mosaicity term is frequency independent and will be only visible in the inhomogeneous linebroadening $\Delta H_0$ and will not affect the determination of $\alpha_{\rm eff}$.

The discussion following the introduction of Eqs.~\ref{2mag} and \ref{alphafit} was focused on crystalline lattice defects as the origin of two-magnon scattering. However any kind on inhomogeneity in the magnetic state of the sample may play the same role. The presence of magnetic dragging, visible for instance for $\phi=30^{\circ}$ in Fig.~\ref{deltah} can create a slight inhomogeneity in the magnetization state for field orientations  close to the hard axis direction and an increase of damping around the hard axis orientation. In any case, this contribution follows also the symmetry of the lattice and it is accounted in the $\Gamma_{\rm 2M}$ parameter. 

Although certain theoretical works point to an anisotropic Gilbert damping in fully epitaxial systems due to its dependence on the density of states at the Fermi energy \cite{gilmore,safonov}, experimentally this has been only seen in ultrathin Fe films \cite{lenz} due to the modification of the electronic structure induced by the interfacial spin-orbit coupling. The anisotropy in $\alpha_{\rm eff}$ presented here can be fully explained by  two-magnon scattering, and therefore an isotropic Gilbert damping can be assumed. 

\section{Conclusions}

Making use of the well defined dependence of the two-magnon scattering mechanism on the in-plane field direction, we have been able to separate this contribution to damping from the isotropic contributions originating from the viscous Gilbert damping mechanism, from spin pumping and from the magnetic proximity effect in Pt. The method can be implemented thanks to the preferential ordering of crystalline defects with respect to the crystallographic directions in epitaxial systems and  therefore cannot be extended to amorphous or polycrystalline magnetic films. This shows the potential of the study of spin pumping related phenomena in ordered systems. Without the contribution of the two-magnon scattering, which depends strongly on the chosen capping layer and defect density, a value of the effective spin mixing conductance $g^{\uparrow\downarrow}_{\rm eff}$ is obtained which is closer to the $g^{\uparrow\downarrow}$ associated only to spin pumping. This approach allows for a better estimation of the spin Hall angle in metals.

\section*{Acknowledgements}

Financial support by  M-era.Net through the HEUMEM project and by the Carl Zeiss Stiftung  is gratefully acknowledged.

\end{document}